\pdfoutput=1
\documentclass[11pt]{article}

\usepackage{EMNLP2023} 
\usepackage{times}
\usepackage{latexsym}
\usepackage{graphicx}
\usepackage{enumitem}
\setlist[itemize]{itemsep=0pt, parsep=0pt, leftmargin = 0pt}
\usepackage{amsthm,amsmath,amssymb}
\usepackage{mathrsfs}
\usepackage{multirow}
\usepackage{enumitem}
\usepackage{array}
\usepackage{latexsym}
\usepackage{graphicx}
\usepackage{subfigure}
\usepackage{enumitem}
\setlist[itemize]{itemsep=0pt, parsep=0pt, leftmargin = 0pt}
\usepackage{amsthm,amsmath,amssymb}
\usepackage{mathrsfs}
\usepackage{booktabs}
\usepackage{float}
\usepackage{booktabs}
\usepackage{multirow}
\usepackage{graphicx}

\usepackage{float}
\usepackage{booktabs}
\usepackage{multirow}

\usepackage[T1]{fontenc}

\usepackage[utf8]{inputenc}

\usepackage{microtype}

\usepackage{inconsolata}

\title{Mitigating Matthew Effect: Multi-Hypergraph Boosted Multi-Interest Self-Supervised Learning for Conversational Recommendation}

\author{{Yongsen Zheng}\textsuperscript{\textnormal{1,2}}, {Ruilin Xu}\textsuperscript{\textnormal{3}}, {Guohua Wang}\textsuperscript{\textnormal{4\textdagger}}, {Liang Lin}\textsuperscript{\textnormal{3,5}}, {Kwok-Yan Lam}\textsuperscript{\textnormal{1,2\textdagger}} \\
\textsuperscript{\textnormal{1}}Nanyang Technological University, Singapore\, \textsuperscript{\textnormal{2}}Digital Trust Centre Singapore\\
 \textsuperscript{\textnormal{3}}Sun Yat-sen University\, \textsuperscript{\textnormal{4}}South China Agricultural University\, \textsuperscript{\textnormal{5}}Peng Cheng Laboratory\\
 \{yongsen.zheng, kwokyan.lam\}@ntu.edu.sg, xurlin5@mail2.sysu.edu.cn\\
 wangguohua@scau.edu.cn, linliang@ieee.org\\
}

\begin{document}
\maketitle

\renewcommand{\thefootnote}{}
\footnotetext{\textsuperscript{\textnormal{\textdagger}}Corresponding author.}

\begin{abstract}
The Matthew effect is a big challenge in Recommender Systems (RSs), where popular items tend to receive increasing attention, while less popular ones are often overlooked, perpetuating existing disparities. Although many existing methods attempt to mitigate Matthew effect in the static or quasi-static recommendation scenarios, such issue will be more pronounced as users engage with the system over time. To this end, we propose a novel framework, Multi-Hypergraph Boosted Multi-Interest Self-Supervised Learning for Conversational Recommendation (HiCore), aiming to address Matthew effect in the Conversational Recommender System (CRS) involving the dynamic user-system feedback loop. It devotes to learn multi-level user interests by building a set of hypergraphs (\emph{i.e.}, item-, entity-, word-oriented multiple-channel hypergraphs) to alleviate the Matthew effec. Extensive experiments on four CRS-based datasets showcase that HiCore attains a new state-of-the-art performance, underscoring its superiority in mitigating the Matthew effect effectively. Our code is available at {\color{blue}https://github.com/zysensmile/HiCore}.
\end{abstract}

\section{Introduction}
Engaging users in ongoing conversations for personalized recommendations, Conversational Recommendation Systems (CRSs) \cite{dialogue_1,dialogue_3} have become a prevalent strategy utilized in diverse fields \cite{e_commerce,music_conversation}. However, CRSs often face a big challenge known as Matthew effect \cite{MatthewEffect1}, captured by the adage "the privileged gain more privilege, while the underprivileged fall further behind." This observation underscores that well-received items or categories in past records garner heightened visibility in future suggestions, whereas less preferred ones frequently face neglect or marginalization.\\
\indent Lately, a multitude of studies have focused on investigating the Matthew effect in relatively unchanging offline recommendation scenarios \cite{MatthewEffect1,MatthewEffect1_1}, identifying two root causes for its occurrence. One cause \cite{MatthewEffect1_3,MatthewEffect1_4,MatthewEffect1_2,MatthewEffect1_1} is the heightened vulnerability of individuals with narrower and uniform preferences or interests to succumb to the pervasive influence of the Matthew effect. This susceptibility often stems from a tendency towards familiarity and comfort, leading to a reinforcement of existing patterns and a limited exploration of diverse alternatives. Another cause \cite{MatthewEffect2_1} arises from the pervasive favoritism towards mainstream items, resulting in a perpetual reinforcement of their prominence, while lesser-known alternatives linger in the shadows. This bias towards popular choices not only perpetuates existing trends but also limits the discoverability of niche or underappreciated options. Thus, the amplification of visibility for widely favored items can overshadow the potential value and diversity offered by less popular but equally deserving alternatives.\\
\indent Despite their effectiveness, most existing methods still suffer from two major limitations. \emph{1) Interactive Strategy}. While many methods have offered valuable insights into the Matthew effect, they often overlook the adverse effects originating from the dynamic user-system feedback loop \cite{feedback_loop}, as they primarily focus on mitigating the Matthew effect in relatively stable offline recommendation settings. In fact, the Matthew effect can intensify as users interact more actively with the system over time, potentially exacerbating concerns such as echo chambers \cite{echo_chamber} and filter bubbles \cite{filter_bubble}. Hence, it is important to address the Matthew effect in the CRS.

\emph{2) Interest Exploration}. Considering that the root cause of the Matthew effect lies in the confinement of user interests \cite{MatthewEffect1_4,MatthewEffect1_3,MatthewEffect1_2,MatthewEffect1_1}, most existing methods focus on leveraging hypergraphs to unveil complex high-order user relationship patterns for exploring user interests. However, these hypergraphs often remain single-channel, constraining their capacity to capture diverse user relation patterns since each hypergraph can only represent a specific type of user patterns. Moreover, these single-channel hypergraphs may risk evolving into traditional Knowledge Graphs (KGs) due to the scarcity of user-item interaction data. Thus, the construction of multi-channel hypergraphs is paramount for exploring multi-level user interests.\\

\indent To address these limitations, we propose the novel framework, Multi-\textbf{H}ypergraph Boosted Multi-\textbf{I}nterest Self-Supervised Learning for \textbf{Co}nversational \textbf{Re}commendation (\textbf{HiCore}), which aims to mitigate the negative impact of Matthew effect when users engage with the system over time in the CRS. It is comprised of Multi-Hypergraph Boosted Multi-Interest Self-Supervised Learning and Interest-Boosted CRS. The former devotes to construct multi-hypergraph (\emph{i.e.}, item-oriented, entity-oriented, and word-oriented triple-channel hypergraphs) to learn multi-level user interests (\emph{i.e.}, item-level, entity-level, word-level triple-channel interests), where triple channels contain the group, joint, and purchase channels. The latter aims to utilize the multi-level interests to enhance both conversation and recommendation tasks when users chat with system over time. Concretely, multi-level user interests are used to effectively generate next utterances in the conversational task, and accurately predict users' interested items in the recommendation task. Extensive experimental results on four benchmarks show that HiCore achieves a new state-of-the-art performance compared all the baselines, and the effectiveness of mitigating Matthew effect in the CRS.\\ 
\indent Overall, our main contributions are included:
\vspace{-8pt}
\begin{itemize}[leftmargin=8pt]
\item To the best of our knowledge, this is the first work to build multi-hypergraph from triple-channel settings for learning multi-interest to mitigate Matthew effect in the CRS.
\item We proposed a novel end-to-end framework HiCore, aiming to use multi-interest enhance both recommendation and conversation tasks.
\item Quantitative and qualitative experimental results show the effectiveness of HiCore and the superiority of HiCore in mitigating Matthew effect. 
\end{itemize}

\section{Related Work}
\subsection{Matthew Effect in Recommendation}
The Matthew effect poses a formidable challenge in recommendation systems. To combat this issue, there are two primary research lines. One line of research focuses on understanding a diverse range of user interests to enhance recommendation diversification \cite{MatthewEffect1_1, MatthewEffect1_2, MatthewEffect1_3, MatthewEffect1_4}. The other line of research \cite{MatthewEffect2_1} is dedicated to mitigating popularity bias to ensure a balanced exposure of items across various categories. For example, Wang et al. \cite{MatthewEffect_QA} conducted a meticulous quantitative analysis, providing valuable insights into the quantitative characteristics of the Matthew effect in collaborative-based recommender systems. Liu et al. \cite{MatthewEffect1} have confirmed the presence and impact of the Matthew effect within the intricate algorithms of YouTube's recommendation system. However, these methods primarily concentrate on exploring the Matthew effect in static recommendation environments, overlooking the crucial interplay of the user-system feedback loop. 

\begin{figure*}[t]
    \centering
    \includegraphics[width=1\textwidth]{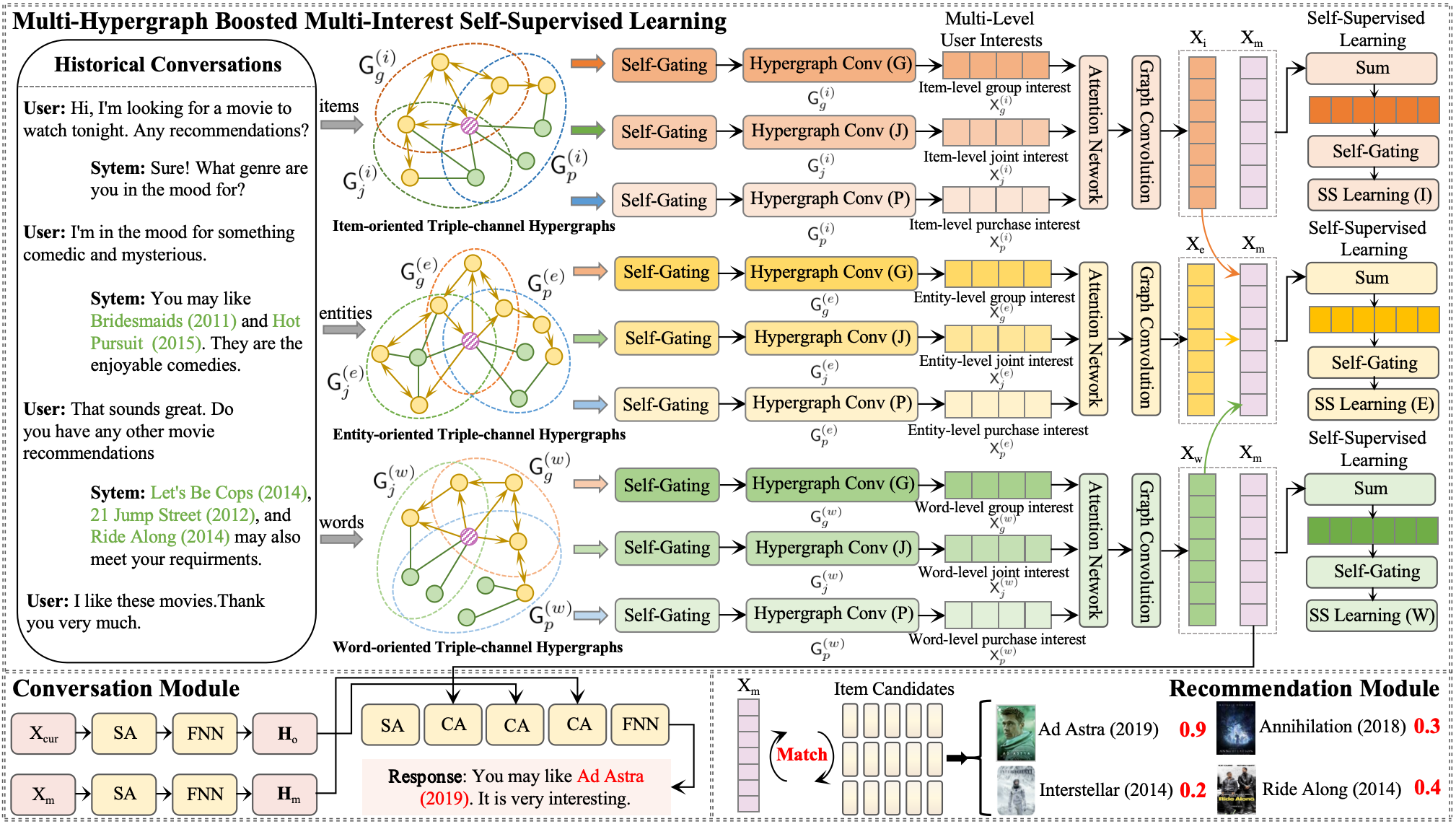}
    \caption{Overview of our HiCore framework. It consists of Multi-Hypergraph Boosted Multi-Interest Self-Supervised Learning and Interested-Boosted CRS. The former aims to learn multi-level user interests, while the latter devotes to generate responses in the conversation module and predict items in the recommendation module.}
    \label{fig:framework}
\end{figure*}

\subsection{Conversational Recommender System}
Unlike traditional static recommendation methods \cite{TRS_1, TRS_2, TRS_new} and interactive recommendation methods \cite{IRS_1, IRS_2}, Conversational Recommendation System \cite{CRS_1, CRS_3, CRS_4, CRS_5} aims to uncover users' genuine intentions and interests through natural language dialogues, thereby offering top-notch recommendations to users. Currently, CRS-based methods can be categorized into two main groups. 1) Attribute-based CRS \cite{deng2021unified,CRS_7,lei2020interactive,ren2021learning,xu2021adapting}, which seeks to delve into user interests by posing queries about items or their attributes. However, this approach primarily relies on predefined templates for response generation, often falling short in producing fluent, human-like natural language expressions. 2) Generated-based CRS \cite{CRS_6,HyCoRec,CRS_8,li2022user,zhou2020improving,C2CRS,shang2023multi}, which can address the shortcomings of attribute-centric CRS by utilizing the Seq2Seq architecture \cite{vaswani2017attention} to integrate a conversation component and a recommendation component, resulting in the creation of smooth and coherent human-like responses. Despite their effectiveness, they face challenges in grasping the varied interests of users because of the restricted and scarce character of user-item interaction data. 

\section{HiCore}\label{sec:HiCore}
Most existing methods \cite{diverse_user_preference_1, diverse_user_preference_2, diverse_user_preference_3} have consistently revealed that individuals with constrained interests are greatly impacted by Matthew effect. Thus, we propose a  novel framework, HiCore, which is comprised of Multi-Hypergraph Boosted Multi-Interest Self-Supervised Learning and Interest-Boosted CRS. The overall pipeline of the proposed HiCore is illustrated in Fig.\ref{fig:framework}.

\subsection{Multi-Hypergraph Boosted Multi-Interest Self-Supervised Learning}\label{sec:Self-Supervised-Learning}
In this section, we will establish multi-hypergraph to learn multi-level user interests to mitigate Matthew effect in the CRS.

\subsubsection{Multi-Hypergraph Boosts Multi-Interest}\label{sec:MHMI}
\begin{figure}[t]
    \centering
    \includegraphics[width=0.5\textwidth]{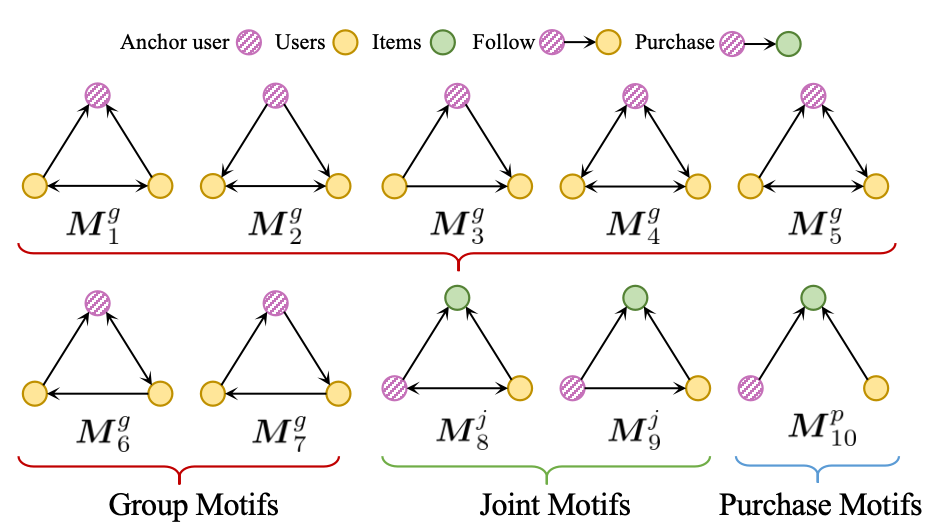}
    \caption{Triangle motifs used in our proposed HiCore.}
    \label{fig:motifs}
\end{figure}

Instead of linking only two nodes per edge as in traditional KGs, hypergraphs extend the notion of edges to connect more than two nodes. By utilizing diverse hypergraphs to encode various high-order user relation patterns, we construct multiple knowledge-oriented triple-channel hypergraphs.\\
\textbf{Item-oriented triple-channel Hypergraphs.} We first build item-oriented hypergraphs from triple channels, \emph{i.e.}, `\emph{Group Channel (g)}', `\emph{Joint Channel (j)}' , and `\emph{Purchase Channel (p)}' via the \emph{Motif} \cite{motif,multichannel}, a commonly utilized tool for capturing complex local structures involving multiple nodes, as illustrated in Fig.\ref{fig:motifs}.

\indent \textbf{Group-channel hypergraph.} Group-channel hypergraphs aim to analyze users' social relations to unveil the dynamics among individuals based on their shared interests, preferences, and characteristics. Understanding group preferences not only consolidates individual tastes but also facilitates collective decisions that benefit the entire group. Formally, we utilize a set of triangular motifs \cite{motif,multichannel} to build the item-oriented group-channel hypergraphs $\textsf{G}^{(i)}_{g}$ as:
\begin{equation}
\begin{aligned}
\textsf{G}^{(i)}_{g}&=(\mathcal{V}^{(i)}_{g}, \mathcal{N}^{(i)}_{g}, \boldsymbol{A}^{(i)}_{M^g_k}).
\end{aligned}
\end{equation}Here $\mathcal{V}^{(i)}_{g}$ represents the set of items derived from the historical conversations, while $\mathcal{N}^{(i)}_{g}=\{\boldsymbol{M}^g_k | 1 \leq k \leq 7\}$ denotes the collection of hyperedges, with each hyperedge representing an occurrence of the specified motif $\boldsymbol{M}^g_k$ in Fig.\ref{fig:motifs}. $\boldsymbol{A}^{(i)}_{M^g_k} \in |\mathcal{V}^{(i)}_{g}| \times |\mathcal{N}^{(i)}_{g}|$ is the group-motif-induced adjacency matrices. Firstly, we need to define the matrix computation of each type of motif. Let $\boldsymbol{H}^{(i)}_k$ be the matrix computation of the motif $\boldsymbol{M}^g_k$, then we can obtain:
\begin{equation}
\left\{
\begin{aligned}
&\boldsymbol{H}^{(i)}_1 = (\boldsymbol{I}^T\boldsymbol{J}) \otimes \boldsymbol{I}^T + (\boldsymbol{J}\boldsymbol{I}) \otimes \boldsymbol{I} +  (\boldsymbol{I}\boldsymbol{I}^T) \otimes \boldsymbol{J},\\
&\boldsymbol{H}^{(i)}_2 = (\boldsymbol{I}\boldsymbol{J}) \otimes \boldsymbol{I} + (\boldsymbol{J}\boldsymbol{I}^T) \otimes \boldsymbol{I}^T + (\boldsymbol{I}^T\boldsymbol{I}) \otimes \boldsymbol{J},\\
&\boldsymbol{H}^{(i)}_3 =  (\boldsymbol{I}\boldsymbol{I}) \otimes \boldsymbol{I} +  (\boldsymbol{I}\boldsymbol{I}^T) \otimes \boldsymbol{I} +  (\boldsymbol{I}^T\boldsymbol{I}) \otimes \boldsymbol{I},\\
&\boldsymbol{H}^{(i)}_4 = (\boldsymbol{J}\boldsymbol{J}) \otimes \boldsymbol{J},\\
&\boldsymbol{H}^{(i)}_5 = (\boldsymbol{J}\boldsymbol{J}) \otimes \boldsymbol{I} + (\boldsymbol{J}\boldsymbol{I}) \otimes \boldsymbol{J} +  (\boldsymbol{I}\cdot \boldsymbol{J}) \otimes \boldsymbol{J},\\
&\boldsymbol{H}^{(i)}_6 = (\boldsymbol{J}\boldsymbol{I}) \otimes \boldsymbol{I}^{T} + (\boldsymbol{I}\boldsymbol{J}) \otimes \boldsymbol{I}^{T} +  (\boldsymbol{I}\boldsymbol{I}) \otimes \boldsymbol{J},\\
&\boldsymbol{H}^{(i)}_7 = (\boldsymbol{I}\boldsymbol{I}) \otimes \boldsymbol{I}^{T},
\end{aligned}
\right.
\label{incidence_matrix} 
\end{equation}where $\otimes$ is the element-wise product, $\boldsymbol{S}$ denote the relation matrix \cite{multichannel}. $\boldsymbol{J}=\boldsymbol{S} \otimes \boldsymbol{S}$, and $\boldsymbol{I} = \boldsymbol{S} - \boldsymbol{J}$ specify the adjacency matrices of the bidirectional and unidirectional social networks (\emph{i.e.}, group motif), respectively. Then, the group-motif-induced adjacency matrices $\boldsymbol{A}^{(i)}_{M^g_k}$ is:
\begin{equation}
\boldsymbol{A}^{(i)}_{M^g_k} = \left\{
\begin{aligned}
& \boldsymbol{H}^{(i)}_1, && {\rm if} \quad \boldsymbol{M}^g_1,\\
&\boldsymbol{H}^{(i)}_2, && {\rm if} \quad \boldsymbol{M}^g_2,\\
& \boldsymbol{H}^{(i)}_3 + (\boldsymbol{H}^{(i)}_5)^T, && {\rm if} \quad \boldsymbol{M}^g_3,\\
&\boldsymbol{H}^{(i)}_4,  && {\rm if} \quad \boldsymbol{M}^g_4,\\
& \boldsymbol{H}^{(i)}_5 + (\boldsymbol{H}^{(i)}_3)^T, && {\rm if} \quad \boldsymbol{M}^g_5,\\
&\boldsymbol{H}^{(i)}_6 +  (\boldsymbol{H}^{(i)}_2)^T, && {\rm if} \quad \boldsymbol{M}^g_6,\\
& \boldsymbol{H}^{(i)}_7 + (\boldsymbol{H}^{(i)}_1)^T, && {\rm if} \quad \boldsymbol{M}^g_7.
\end{aligned}
\right.
\label{adjecent_matrix_group_channel}
\end{equation}
If \((\boldsymbol{A}^{(i)}_{M^g_k})_{n,r} = 1\), it signifies that the node \(n\) and the node \(r\) co-occur in a single instance of \(\boldsymbol{M}^g_k\). When two nodes appear in multiple instances, it turns to be \((\boldsymbol{A}^{(i)}_{M^g_k})_{n,r} = \#(\text{n, r occur in the same instance of } \boldsymbol{M}^g_k)\).

\indent \textbf{Joint-channel hypergraph.} The joint channel reflects the scenario of shared behaviors among friends in a social network. When friends purchase the same items, it not only suggests similarities in tastes and interests but also hints at deeper levels of interaction and trust. This phenomenon of "friends purchasing the same item" may facilitate information dissemination and interaction within the social network, strengthening social relationships, and to some extent, reflecting influence and collective behavior within the social network. 
Therefore, by identifying and analyzing the joint motifs, the item-oriented joint-channel hypergraph $\textsf{G}^{(i)}_{j}$ is:
\begin{equation} 
\left\{
\begin{aligned}
&\textsf{G}^{(i)}_{j}=(\mathcal{V}^{(i)}_{j}, \mathcal{N}^{(i)}_{j}, \boldsymbol{A}^{(i)}_{M^j_k}),\\
&\boldsymbol{H}^{(i)}_8 = (\boldsymbol{R}\boldsymbol{R}^T) \otimes \boldsymbol{J},\\
&\boldsymbol{H}^{(i)}_9 = (\boldsymbol{R}\boldsymbol{R}^T) \otimes \boldsymbol{I},\\
&\boldsymbol{A}^{(i)}_{M^j_k} = \boldsymbol{H}^{(i)}_8, && {\rm if} \quad \boldsymbol{M}^j_8,\\
&\boldsymbol{A}^{(i)}_{M^j_k} =\boldsymbol{H}^{(i)}_9 + [\boldsymbol{H}^{(i)}_9]^T, && {\rm if} \quad \boldsymbol{M}^j_9,\\
\end{aligned}
\right.
\label{adjecent_matrix_joint_channel}
\end{equation}where $\mathcal{V}^{(i)}_{j}$, and $\mathcal{N}^{(i)}_{j}=\{{M^j_k} | 8 \leq k \leq 9\}$ denote the item set, and the hyperedge set, respectively. Each hyperedge is induced from each type of joint motif, depicted in Fig.\ref{fig:motifs}. \( \boldsymbol{R} \) is a binary matrix that records user-item interactions, and \( \boldsymbol{A}^{(i)}_{M^j_k} \) denotes the joint-motif-induced adjacency matrices.

\indent \textbf{Purchase-channel hypergraph.} Additionally, we should also take into account users who do not have explicit social connections. Therefore, the analysis is non-exclusive and delineates the implicit higher-order social relationships among users who lack direct social ties but still purchase the same items. By considering these users without overt social links, we can uncover hidden patterns of social influence and affiliation that transcend traditional network structures. Thus, the item-oriented purchase-channel hypergraph $\textsf{G}^{(i)}_{p}$ can be induced from the purchase motif $\boldsymbol{M}^{p}_{10}$ as follows:
\begin{equation}
\begin{aligned}
\textsf{G}^{(i)}_{p}&=(\mathcal{V}^{(i)}_{p}, \mathcal{N}^{(i)}_{p}, \boldsymbol{A}^{(i)}_{M^p_k}),\\
\boldsymbol{A}^{(i)}_{M^p_k} &= \boldsymbol{H}^{(i)}_{10} = \boldsymbol{R}\boldsymbol{R}^T, \quad {\rm if} \quad \boldsymbol{M}^p_{10},\\
\end{aligned}
\label{adjecent_matrix_purchase_channel}
\end{equation}here $\mathcal{V}^{(i)}_{p}$ and $\mathcal{N}^{(i)}_{p} = \{{M^p_k} | k=10 \}$ are the item set and hyperedge set, respectively. Specifically, the hyperedge set, depicted in Fig.\ref{fig:motifs}. $\boldsymbol{A}^{(i)}_{M^p_k}$ is the purchase-motif-induced adjacency matrices.

\noindent \textbf{Entity-oriented triple-channel hypergraphs.} To tackle the sparsity and constraints inherent in historical user-item interaction data, we leverage the rich DBpedia KG \cite{DBpedia} to build an entity-oriented hypergraph. More precisely, we identify individual items referenced in historical conversations as entities and their $k$-hop neighbors to construct each hyperedge. This method enables us to capture shared semantic nuances among the broader network of neighbors. Similar to item-oriented triple-channel hypergraphs, we build the entity-oriented hypergraphs from triple channel setting. Formally, the entity-oriented hypergraphs $\textsf{G}^{(e)}_{c}$ from triple-channel $c$ can be given as:
\begin{equation}
\begin{aligned}
\textsf{G}^{(e)}_{c}=(\mathcal{V}^{(e)}_{c}, \mathcal{N}^{(e)}_{c}, \boldsymbol{A}^{(e)}_{M^c_k}).\\
\end{aligned}
\end{equation}Here $c \in \{g, j, p\}$ represents triple channels (\emph{i.e.}, group, joint, and purchase channel). $\mathcal{V}^{(e)}_{c}$ denotes the entities from triple-channel setting. These entities are  $k$-hop neighbors extracted from the historical conversations. $\mathcal{N}^{(e)}_{c}$ means the hyperedge set induced from different motifs. Each hyperedge is an instance of the Motif. $\boldsymbol{A}^{(e)}_{M^c_k}$ represents the group-channel, joint-channel, and purchase-channel adjacency matrices, they are defined as Eq.(\ref{adjecent_matrix_group_channel}), Eq.(\ref{adjecent_matrix_joint_channel}), and Eq.(\ref{adjecent_matrix_purchase_channel}), respectively.

\noindent \textbf{Word-oriented triple-channel hypergraphs.} The significance of keywords exchanged during conversations is paramount in grasping users' requirements. By scrutinizing notable words, we can pinpoint specific inclinations, a critical aspect in modeling an array of user tastes. To realize this objective, we construct a lexeme-centric hypergraph utilizing the lexicon-focused ConceptNet \cite{ConceptNet} KG to unveil semantic associations such as synonymy, antonyms, and co-occurrence. Based on these analysis, the word-oriented hypergraphs from group-, joint-, and purchase-channel can be expressed as:
\begin{equation}
\begin{aligned}
\textsf{G}^{(w)}_{c}&=(\mathcal{V}^{(w)}_{c}, \mathcal{N}^{(w)}_{c}, \boldsymbol{A}^{(w)}_{M^c_k}),
\end{aligned}
\end{equation}where $\mathcal{V}^{(w)}_{c}$ is the words from $k$-hop neighbors. $\mathcal{N}^{(w)}_{c}$ denotes the hyperedge set from different motifs, including group, joint, purchase motifs. $\boldsymbol{A}^{(w)}_{M^c_k}$ are the word-oriented adjacency matrices induced from triple channels, illustrated in Eq.(\ref{adjecent_matrix_group_channel}) $\sim$ Eq.(\ref{adjecent_matrix_purchase_channel}). 

\subsubsection{Multi-Interest Self-Supervised Learning}\label{sec:MISSL}
After constructing a series of hypergraphs from triple-channel setting, we will construct multi-level user interests via the hypergraph convolution network \cite{multichannel}, which can be written as:
\begin{equation}
\begin{aligned}
\boldsymbol{P}^{(l+1)}_c =  \boldsymbol{D}^{-1}_{c} \boldsymbol{K}_{c} \boldsymbol{L}^{-1}_{c} \boldsymbol{K}^T_{c} \boldsymbol{P}^{(l)}_{c} =\boldsymbol{\widehat{D}}^{-1}_{c} \boldsymbol{A}^{(i)}_c \boldsymbol{P}^{(l)}_c,
\end{aligned}
\label{base_user_embedding}
\end{equation}where $\boldsymbol{P}^{(l)}_c$ and $\boldsymbol{P}^{(l+1)}_c$ represent the output of the $l$-th and $(l+1)$-th layers, respectively. Specifically, the base user embedding is $\boldsymbol{P}^{(0)}_c = f^c_{\rm gate} (\boldsymbol{P}^{(0)})$, and $f^c_{\rm gate}(\cdot)$ is self-gating units (SGUs) to control the information flow to different channel from the base user embedding $\boldsymbol{P}^{(0)}$. $\boldsymbol{D}_{c}$ is the degree matrix of $\boldsymbol{A}_c$, which is the summation of the motifs without considering self-connections \cite{multichannel}. 
In terms of the group motifs, it can be defined as $\boldsymbol{A}^{(i)}_c=\sum^7_{k=1} \boldsymbol{A}^{(i)}_{M_k}$, in terms of joint motifs, $\boldsymbol{A}^{(i)}_c=\boldsymbol{A}^{(i)}_{M_8} + \boldsymbol{A}^{(i)}_{M_9}$, and from the point of the purchase motifs, $\boldsymbol{A}^{(i)}_c=\boldsymbol{A}^{(i)}_{M_{10}} - (\boldsymbol{A}^{(i)}_{M_8} + \boldsymbol{A}^{(i)}_{M_9})$. Based on these analysis, the item-level interests from triple channel setting (\emph{i.e.}, $\boldsymbol{X}^{(i)}_{g}$, $\boldsymbol{X}^{(i)}_{j}$, $\boldsymbol{X}^{(i)}_{p}$), the entity-level interests from triple channel setting (\emph{i.e.}, $\boldsymbol{X}^{(e)}_{g}$, $\boldsymbol{X}^{(e)}_{j}$, $\boldsymbol{X}^{(e)}_{p}$), the word-level interests from triple channel setting (\emph{i.e.}, $\boldsymbol{X}^{(w)}_{g}$, $\boldsymbol{X}^{(w)}_{j}$, $\boldsymbol{X}^{(w)}_{p}$) can be defined as:
\begin{equation}
\begin{aligned}
\boldsymbol{X}^{(h)}_{g} &= \boldsymbol{\widehat{D}}^{-1}_{g} (\sum^7_{k=1} \boldsymbol{A}^{(h)}_{M^g_k}) \boldsymbol{P}^{(L)}_g;\\
\boldsymbol{X}^{(h)}_{j} &= \boldsymbol{\widehat{D}}^{-1}_{j} (\boldsymbol{A}^{(h)}_{M^j_8} + \boldsymbol{A}^{(h)}_{M^j_9}) \boldsymbol{P}^{(L)}_j;\\
\boldsymbol{X}^{(h)}_{p} &= \boldsymbol{\widehat{D}}^{-1}_{p} (\boldsymbol{A}^{(h)}_{M^p_{10}} - (\boldsymbol{A}^{(h)}_{M^j_8} + \boldsymbol{A}^{(h)}_{M^j_9}))\boldsymbol{P}^{(L)}_p.\\
\end{aligned}
\label{diverse_interests}
\end{equation}Here $h \in \{i, e, w\}$, and $L$ is the last hypergraph convolution layer. Then, we adopt the attention network $\textsf{Atta}(\cdot)$ and graph convolution $\textsf{GConv}(\cdot)$ to learn final multi-interest $\boldsymbol{X}_{m}$ as:
\begin{equation}
\begin{aligned}
&\boldsymbol{X}_{i} = \textsf{GConv}(\textsf{Atta}(\boldsymbol{X}^{(i)}_{g};\boldsymbol{X}^{(i)}_{j};\boldsymbol{X}^{(i)}_{p})),\\
&\boldsymbol{X}_{e} =\textsf{GConv}(\textsf{Atta}(\boldsymbol{X}^{(e)}_{ g};\boldsymbol{X}^{(e)}_{j};\boldsymbol{X}^{(e)}_{p})),\\
&\boldsymbol{X}_{w} =\textsf{GConv}(\textsf{Atta}\boldsymbol{X}^{(w)}_{ g};\boldsymbol{X}^{(w)}_{j};\boldsymbol{X}^{(w)}_{p})),\\
&\boldsymbol{X}_{m} = \textsf{Atta}(\boldsymbol{X}_i;\boldsymbol{X}_e;\boldsymbol{X}_w).
\end{aligned}
\end{equation}Here \(;\) is the concatenation operation. Finally, we use InfoNCE \cite{multichannel} as our learning objective to conduct the self-supervised learning as:
\begin{equation}
\begin{aligned}
\mathcal{L}_s = &- \sum_{h} \Big \{ \sum_{u \in U} {\rm log} \sigma(f (\boldsymbol{X}_{m}), z^h_u) - f(\boldsymbol{X}_{m}, \hat{z}^h_u))\\
& + \sum_{u \in U} {\rm log} \sigma(f(z^h_u, \textbf{k}^h) - f(\hat{z}^h_u, \textbf{k}^h) \Big \}.
\end{aligned}
\end{equation} Here $z^h_u=f^h_{\rm gate}(f_s(\boldsymbol{X}_{h};\textbf{p}^h_u)$, $f_s(\cdot)$ is the sum operation, $\hat{z}^h_u$ is the negative example by shuffling both rows and columns of $z^h_u$, and $h$ is defined as Eq.(\ref{diverse_interests}). \(f(\cdot) \in \mathbb{R}^{d \times d}\) serves as the discriminator, evaluating the alignment between two input vectors. Specifically, \(\textbf{k}^h = f_{\rm out}(Z_h)\), where \(Z_h\) and \(\boldsymbol{X}_{m}\) are ground truths for each other, and \(f_{\rm out}(\cdot)\) aims to perform permutation invariant \cite{multichannel}.

\subsection{Interest-Boosted CRS}\label{sec:Interest-Boosted-CRS}
To mitigate Matthew effect in the CRS, we employ multi-interest \(\boldsymbol{X}_{m}\) to enhance both recommendation and conversation tasks.

\subsubsection{Recommendation Module}
Recommendation module is to precisely forecast items for users via dynamic natural conversations. To improve recommendation diversity, we use \(\boldsymbol{X}_{m}\) to select target items as $\textsf{P}_{\rm rec} =  \boldsymbol{X}_{m} \times \boldsymbol{V}_{\rm cand}$, where $\boldsymbol{V}_{\rm cand}$ is the embeddings of all candidate items. Finally, we adopt cross-entropy loss \cite{shang2023multi} to learn the recommendation task:
\begin{equation}
\begin{aligned}
\mathcal{L}_{\rm rec} = - \sum^{B}_{b=1} \sum^{|\mathcal{I}|}_{a=1} & \Big \{ -(1-\textsf{Y}_{ab}) \cdot {\rm log}(1-{\textsf{P}}^{(b)}_{\rm rec}(a))\\
&+ \textsf{Y}_{ab} \cdot {\rm log}({\textsf{P}}^{(b)}_{\rm rec}(a)) \Big \},
\end{aligned}
\end{equation}where $\textsf{Y}_{ab} \in \{0, 1\}$, $B$, and $|\mathcal{I}|$ are the target label, mini-batch size, the size of item set, respectively.

\subsubsection{Conversation Module}
Conversation module centers on crafting appropriate dialogue responses. Next, we use multi-interest $\boldsymbol{X}_{m}$ to fed into Transformer $\textsf{MHA}(\cdot)$ to produce informative responses. Suppose $\boldsymbol{\rm Y}^{n-1}$ is the output of the last time unit, then the current one $\boldsymbol{\rm Y}^{n}$ is:
\begin{equation}
\begin{aligned}
\boldsymbol{\rm A}^{n}_0 &= \textsf{MHA}(\boldsymbol{\rm Y}^{n-1}, \boldsymbol{\rm Y}^{n-1}, \boldsymbol{\rm Y}^{n-1}),\\
\boldsymbol{\rm A}^{n}_1 &= \textsf{MHA}(\boldsymbol{\rm A}^{n}_0, \boldsymbol{X}_{m}, \boldsymbol{X}_{m}),\\
\boldsymbol{\rm A}^{n}_2 &= \textsf{MHA}(\boldsymbol{\rm A}^{n}_1,  \boldsymbol{X}_{\rm cur}, \boldsymbol{X}_{\rm cur}),\\
\boldsymbol{\rm A}^{n}_3 &= \textsf{MHA}(\boldsymbol{\rm A}^{n}_1, \boldsymbol{X}_{\rm his}, \boldsymbol{X}_{\rm his}),\\
\boldsymbol{\rm A}^{n}_4 &= \beta \cdot \boldsymbol{\rm A}^{n}_2 + (1 - \beta) \cdot \boldsymbol{\rm A}^{n}_3,\\
 \boldsymbol{\rm Y}^{n} &= \textsf{FFN}(\boldsymbol{\rm A}^{n}_4),
\label{response_generator}
\end{aligned}
\end{equation}where $\boldsymbol{X}_{\rm cur}$ and $\boldsymbol{X}_{\rm his}$ are the current and historical conversations, respectively. $\beta$ is hyper-parameters to control the information flow. Then, we use cross-entropy loss to learn the conversation task:
\begin{equation}
\begin{aligned}
\mathcal{L}_{\rm conv} = -\sum_{b=1}^B \sum_{t=1}^T {\rm log}(\textsf{P}_{\rm conv}( s_t | \{s_{t-1}\})),\\
\textsf{P}_{\rm conv}(\cdot) = p_1(s_t|Y_i) + p_2(s_t|\textsf{P}_{\rm rec})+ p_3(s_t|\textsf{P}_{\rm rec}),\\
\label{conversation_loss_2}
\end{aligned}
\end{equation}where \(\textsf{P}_{\rm conv}(\cdot) \) is the probability of the next token when given a sequence \(\{s_{t-1}\}=s_1, s_2, \cdots, s_{t-1}\), where \(s_t\) signifies the \(t\)-th utterance. \(p_1(\cdot)\), \(p_2(\cdot)\), and \(p_3(\cdot)\) denote the vocabulary probability, vocabulary bias, and copy probability, respectively. \(T\) is the truncated length of utterances.

\subsection{Challenges Discussion}\label{sec:discussion}
Throughout the developmental journey of hypergraphs, we surmounted several significant challenges, elaborated upon below:\\
\indent \emph{1)} Hypergraph Construction Challenge: During the project's initial stages, the real-time construction of hypergraphs presented a bottleneck, resulting in delays. Through the strategic repositioning of this operation to the data preprocessing phase, we adeptly extracted essential subgraphs, leading to a noteworthy reduction in training time. This adjustment enhanced efficiency, streamlined processes, and improved performance.\\
\indent \emph{2)} Graph Storage Challenge: The transition to sparse graph storage mechanisms is pivotal in enhancing efficiency, streamlining computation time, and optimizing memory utilization. Embracing this shift not only boosts the system's performance but also establishes a robust foundation for scalable and resource-efficient operations.\\
\indent \emph{3)} Model Training Challenge: With the emergence of a series of hypergraphs, optimizing the efficiency of model training becomes paramount. Consequently, we redefined our strategy by dispersing hypergraphs across multiple computing cards, enabling parallel computation and achieving a significant boost in the model's runtime speed.

\begin{table*}
\small
\setlength{\tabcolsep}{8mm}
\setlength{\abovecaptionskip}{4pt}  
\centering
\renewcommand{\arraystretch}{1.1}
\begin{tabular}{l@{\hskip 0.00in}
l@{\hskip 0.13in}
c@{\hskip 0.08in}c@{\hskip 0.08in}c@{\hskip 0.08in}c@{\hskip 0.08in}c@{\hskip 0.08in}c@{\hskip 0in}
c@{\hskip 0.18in}
c@{\hskip 0.08in}c@{\hskip 0.08in}c@{\hskip 0.08in}c@{\hskip 0.08in}c@{\hskip 0.08in}c@{\hskip 0in}
c@{\hskip 0.13in}}
\toprule
\multirow{2}{*}{\textbf{}}&
\multirow{2}{*}{\textbf{Model}}&
\multicolumn{6}{c}{REDIAL}&
&
\multicolumn{6}{c}{TG-REDIAL}
&\\
\cline{3-8}
\cline{10-15}
\rule{0pt}{10pt}
&&R@10&R@50&M@10&M@50&N@10&N@50&&R@10&R@50&M@10&M@50&N@10&N@50\\
&TextCNN&0.0644&0.1821&0.0235&0.0285&0.0328&0.0580&
&0.0097&0.0208&0.0040&0.0045&0.0053&0.0077\\
&SASRec&0.1117&0.2329&0.0540&0.0593&0.0674&0.0936&
&0.0043&0.0178&0.0011&0.0017&0.0019&0.0047\\
&BERT4Rec&0.1285&0.3032&0.0475&0.0555&0.0663&0.1045&
&0.0043&0.0226&0.0013&0.0020&0.0020&0.0058\\
&KGSF&0.1785&0.3690&0.0705&0.0796&0.0956&0.1379&
&0.0215&0.0643&0.0069&0.0087&0.0103&0.0194\\
&TG-ReDial&0.1679&0.3327&0.0694&0.0771&0.0924&0.1286&
&0.0110&0.0174&0.0048&0.0050&0.0062&0.0076\\
&ReDial&0.1705&0.3077&0.0677&0.0738&0.0925&0.1222&
&0.0038&0.0165&0.0012&0.0017&0.0018&0.0045\\
&KBRD&0.1796&0.3421&0.0722&0.0800&0.0972&0.1333&
&0.0201&0.0501&0.0077&0.0090&0.0106&0.0171\\
&BART&0.1693&0.3783&0.0646&0.0744&0.0888&0.1350&
&0.0047&0.0187&0.0012&0.0017&0.0020&0.0048\\
&BERT&0.1608&0.3525&0.0597&0.0688&0.0831&0.1255&
&0.0040&0.0194&0.0011&0.0017&0.0018&0.0050\\
&XLNet&0.1569&0.3590&0.0583&0.0677&0.0811&0.1255&
&0.0040&0.0187&0.0011&0.0017&0.0017&0.0048\\
&KGConvRec&0.1819&0.3587&0.0711&0.0794&0.0969&0.1358&
&0.0220&0.0524&0.0088&0.0102&0.0119&0.0185\\
&MHIM&0.1966&0.3832&0.0742&0.0830&0.1027&0.1440&
&0.0300&0.0783&0.0108&0.0129&0.0152&0.0256\\
&\textbf{HiCore*}&\textbf{0.2192}&\textbf{0.4163}&\textbf{0.0775}&\textbf{0.0874}&\textbf{0.1107}&\textbf{0.1558}&
&\textbf{0.0270}&\textbf{0.0769}&\textbf{0.0880}&\textbf{0.1074}&\textbf{0.0152}&\textbf{0.0225}\\
\bottomrule
\end{tabular}
\caption{\label{tab:recommendation} Recommendation results on REDIAL and TG-REDIAL datasets. * indicates statistically significant improvement (\emph{p} < 0.05) over all baselines.}
\end{table*}

\section{Experiments and Analyses}
To fully evaluate the proposed HiCore, we conduct experiments to answer the following questions:
\vspace{-8pt}
\begin{itemize}
\item \textbf{RQ1:} How does HiCore perform compared with all baselines in the recommendation task?
\item \textbf{RQ2:} How does HiCore perform compared with all baselines in the conversation task?
\item \textbf{RQ3:} How does HiCore mitigate Matthew effect in the CRS?
\item \textbf{RQ4:} How do parameters affect our HiCore?
\item \textbf{RQ5:} How do different hypergraphs contribute to the performance?
\end{itemize}

\subsection{Experimental Protocol}
\textbf{Datasets.} We assess the effectiveness of our proposed HiCore through comprehensive evaluations on four CRS-based benchmarks: REDIAL \cite{TDCR}, TG-REDIAL \cite{Topic-Guided}, OpenDialKG \cite{opendialkg}, and DuRecDial \cite{durecdial}. The REDIAL dataset comprises 11,348 dialogues involving 956 users and 6,924 items, while the TG-REDIAL dataset encompasses 10,000 dialogues with 1,482 users and 33,834 items. To provide a holistic evaluation of our proposed methodology, we integrate two cross-domain datasets, OpenDialKG and DuRecDial, which cover a wide array of domains including movies, music, books, sports, restaurants, news, and culinary experiences.

\noindent \textbf{Baselines.} We compared our HiCore with the following state-of-the-art methods \textbf{TextCNN} \cite{TextCNN}, \textbf{SASRec} \cite{SASRec}, \textbf{BERT4Rec} \cite{BERT4Rec}, \textbf{KBRD} \cite{chen2019towards}, \textbf{Trans.} \cite{Trans}, \textbf{ReDial} \cite{li2018towards}, \textbf{KGSF} \cite{zhou2020improving}, \textbf{KGConvRec} \cite{sarkar2020suggest}, \textbf{XLNet} \cite{XLNet}, \textbf{BART} \cite{BART}, \textbf{BERT} \cite{BERT}, \textbf{DialoGPT} \cite{DialoGPT}, \textbf{UniCRS} \cite{UniCRS}, \textbf{GPT-3} \cite{GPT3}, \textbf{C2-CRS} \cite{C2CRS}, \textbf{LOT-CRS} \cite{LOTCRS}, \textbf{MHIM} \cite{shang2023multi}, and \textbf{HyCoRec} \cite{HyCoRec}.

\subsection{Recommendation Performance (RQ1)}
In accordance with \cite{shang2023multi}, we utilize Recall@\emph{K} (R@\emph{K}), MRR@\emph{K} (M@\emph{K}), and NDCG@\emph{K} (N@\emph{K}) (\emph{K}=1, 10, 50) to assess the recommendation task. Analyzing the results presented in Table \ref{tab:recommendation} and Table  \ref{tab:Recommendation_opendialkg_durecdial}, it is evident that our proposed method, HiCore, consistently outperforms all the comparison baselines.

\begin{table}
\small
\setlength{\tabcolsep}{5mm}
\setlength{\abovecaptionskip}{6pt}  
\centering
\renewcommand{\arraystretch}{1.1}
\begin{tabular}{l@{\hskip 0.1in}
c@{\hskip 0.05in}c@{\hskip 0.05in}
c@{\hskip 0.15in}
c@{\hskip 0.05in}c@{\hskip 0.05in}
c@{\hskip 0.1in}}
\hline
\multirow{2}{*}{\textbf{Model}}&
\multicolumn{2}{c}{OpenDialKG}&
&
\multicolumn{2}{c}{DuRecDial}
&\\
\cline{2-3}
\cline{5-6}
\rule{0pt}{10pt}
&R@1&R@10&
&R@1&R@10&\\
\hline
KBRD&0.1448&0.3162&
&0.0618&0.3971&\\
KGSF&0.0626&0.1757&
&0.1395&0.4367&\\
ReDial&0.0008&0.0134&
&0.0005&0.0336&\\
TGReDial&0.2149&0.4035&
&0.0956&0.4882&\\
HyCoRec&0.2742&0.4490&
&0.1279&0.4750&\\
\textbf{HiCore*}&\textbf{0.2628}&\textbf{0.4526}&
&\textbf{0.1735}&\textbf{0.5471}\\
\hline
\multirow{2}{*}{\textbf{Model}}&
\multicolumn{2}{c}{OpenDialKG}&
&
\multicolumn{2}{c}{DuRecDial}
&\\
\cline{2-3}
\cline{5-6}
\rule{0pt}{10pt}
&Dist-2&Dist-3&
&Dist-2&Dist-3&\\
\hline
KBRD&0.3192&1.7660&
&0.5180&1.5500&\\
KGSF&0.1687&0.5387&
&0.1389&0.3862&\\
ReDial&0.1579&0.5808&
&0.1095&0.3981&\\
TGReDial&0.4836&2.1430&
&0.5453&2.0030&\\
HyCoRec&2.8190&4.7710&
&1.0820&2.4440&\\
\textbf{HiCore*}&\textbf{2.8430}&\textbf{4.8120}&
&\textbf{1.0940}&\textbf{2.4280}\\
\hline
\end{tabular}
\caption{\label{tab:Recommendation_opendialkg_durecdial} {\color{black}Results on both recommendation and conversation tasks on OpenDialKG and DuRecDial datasets involving various domains. * indicates statistically significant improvement (\emph{p} < 0.05) over all baselines.}}
\end{table}

There exist multiple crucial facets contributing to the advancement of our proposed HiCore method: (a) Diversification of hypergraphs: we introduced a diverse set of hypergraphs, including item-oriented, entity-oriented, and word-oriented hypergraphs. This expansion aims to go beyond the traditional pairwise interactions, broadening the scope of user interest modeling by incorporating interactions among multiple nodes. (b) Exploration of hypergraph configurations: moving beyond the conventional triple-channel model, we delved into various hypergraph configurations like group-channel, joint-channel, and purchase-channel. These configurations were designed to cater not only to social connections but also individual preferences, enhancing the system's adaptability. (c) Integration of multi-level user interests: transitioning from the triple-channel structure, we integrated these hypergraphs to capture multi-level user interests. This strategic shift helps alleviate the Matthew effect in the CRS involving the dynamic user-system feedback loop. This comprehensive approach highlights the innovation and adaptability of HiCore in addressing the intricacies of user interest modeling and enhancing recommendation system performance.

\begin{table}
\small
\setlength{\tabcolsep}{0.1mm}
\setlength{\abovecaptionskip}{2pt}  
\centering
\renewcommand{\arraystretch}{1.1}
\begin{tabular}{l@{\hskip 0.1in}
c@{\hskip 0.05in}c@{\hskip 0.05in}c@{\hskip 0in}
c@{\hskip 0.15in}
c@{\hskip 0.05in}c@{\hskip 0.05in}c@{\hskip 0in}
c@{\hskip 0.1in}}
\hline
\multirow{2}{*}{\textbf{Model}}&
\multicolumn{3}{c}{REDIAL}&
&
\multicolumn{3}{c}{TG-REDIAL}
&\\
\cline{2-4}
\cline{6-8}
\rule{0pt}{10pt}
&Dist-2&Dist-3&Dist-4&
&Dist-2&Dist-3&Dist-4&\\
\hline
ReDial&0.0214&0.0659&0.1333&
&0.2178&0.5136&0.7960&\\
Trans. &0.0538&0.1574&0.2696&
&0.2362&0.7063&1.1800&\\
KGSF&0.0572&0.2483&0.4349&
&0.3891&0.8868&1.3337&\\
KBRD&0.0765&0.3344&0.6100&
&0.8013&1.7840&2.5977&\\
DialoGPT&0.3542&0.6209&0.9482&
&1.1881&2.4269&3.9824&\\
GPT-3&0.3604&0.6399&0.9511&
&1.2255&2.5713&4.0713&\\
UniCRS&0.2464&0.4273&0.5290&
&0.6252&2.2352&2.5194&\\
C2-CRS&0.2623&0.3891&0.6202&
&0.5235&1.9961&2.9236&\\
LOT-CRS&0.3312&0.6155&0.9248&
&0.9287&2.4880&3.4972&\\
MHIM&0.3278&0.6204&0.9629&
&1.1100&2.3520&3.8200&\\
\textbf{HiCore*}&\textbf{0.5871}&\textbf{1.1170}&\textbf{1.7500}&
&\textbf{2.8610}&\textbf{5.7440}&\textbf{8.4160}\\
\hline
\end{tabular}
\caption{\label{tab:conversation} Conversation results on REDIAL and TG-REDIAL datasets. * indicates statistically significant improvement (\emph{p} < 0.05) over all baselines.}
\end{table}

\subsection{Conversational Performance (RQ2)}
For the conversation task, we use Distinct \emph{n}-gram (Dist-\emph{n}) \cite{shang2023multi} (\emph{n}=2,3,4) to estimate the conversation task. Table \ref{tab:Recommendation_opendialkg_durecdial} and Table \ref{tab:conversation} indicate a significant performance superiority of our HiCore. For example, HiCore gains 123.83\%, 138.27\%, 77.26\%, 65.75\%, 62.90\%, and 79.10\% improvements on Dist-2 against the strong baselines including, C2-CRS, UniCRS, LOT-CRS, DialoGPT, GPT-3, and MHIM on the REDIAL dataset, respectively. It also gains 446.51\%, 357.61\%, 208.07\%, 140.80\%, 133.46\%, and 157.75\% improvements on Dist-2 against the strong baselines including, C2-CRS, UniCRS, LOT-CRS, DialoGPT, GPT-3, and MHIM on the REDIAL dataset, respectively. 

The improvement in HiCore can be attributed to the following reasons: (a) Our HiCore focuses on constructing a diverse set of hypergraphs, encompassing item-oriented, entity-oriented, and word-oriented triple-channel hypergraphs. These structures effectively capture intricate local patterns through motif analysis, enabling the exploration of high-order user behaviors. This proves invaluable in generating informative and high-quality response utterances. (b) HiCore is dedicated to mitigating the Matthew effect that may occur as users engage with the system over time. By learning multi-level user interests from the hypergraphs, the system can adapt to users' evolving preferences. This strategic approach enables the CRS to provide a varied array of responses that align with the diverse interests of the users.

\subsection{Study on Matthew Effect (RQ3)}
Given our goal of mitigating the Matthew effect that may arise as users interact with the system over time, we engage in a series of experiments comparing the proposed method with the most robust baselines. This investigation seeks to determine the efficacy of HiCore in effectively alleviating the Matthew effect. Considering the key strategy to mitigate Matthew effect is to improve the recommendation diversification, and thus we use the  diversify-based evaluation metrics \emph{Coverage}@k (C@k), \emph{Average Popularity} (A@K) of Recommended Items and \emph{Long Tail Recommendation Ratio} (L@K) to comprehensively evaluate the efficacy of our proposed method in mitigating the Matthew Effect.

\begin{figure}[t]
    \centering
    \includegraphics[width=0.49\textwidth]{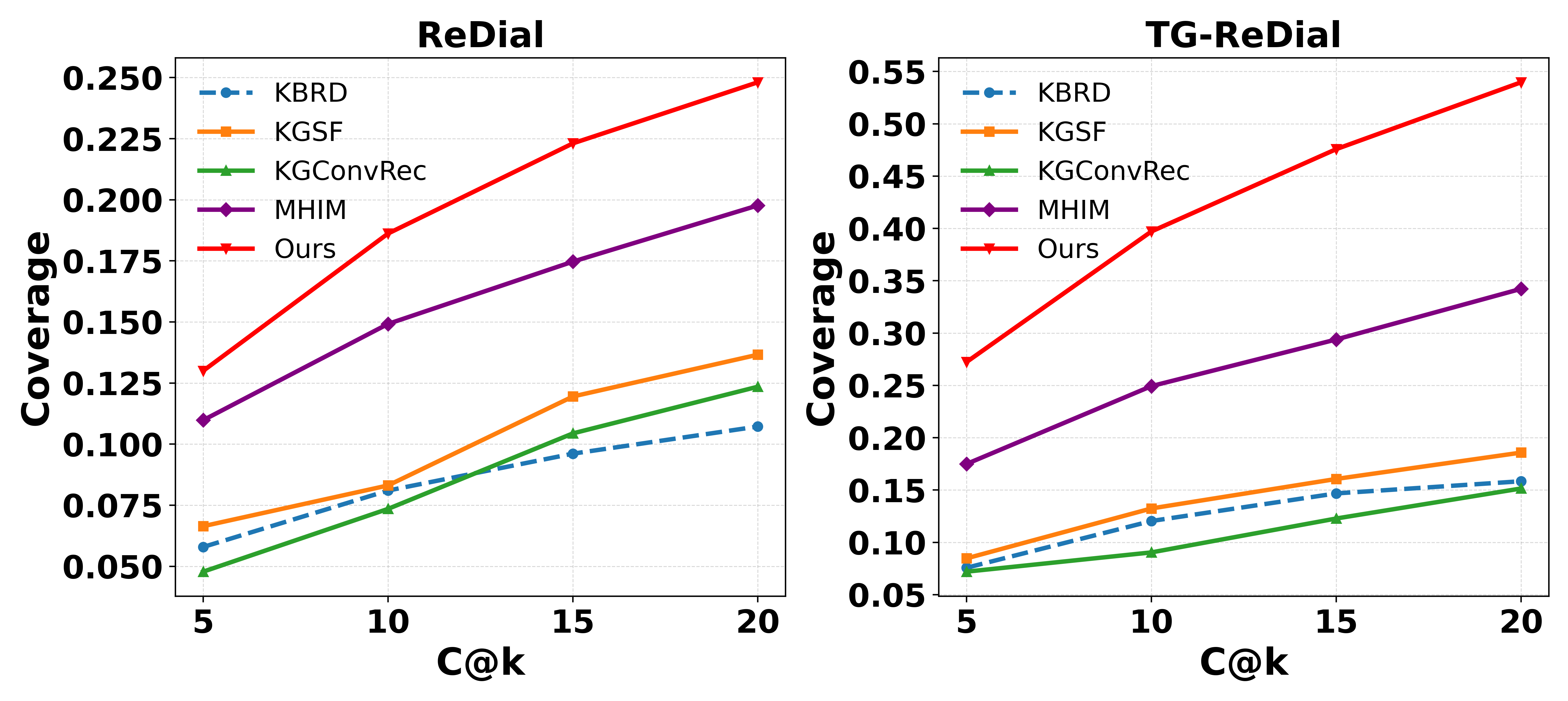}
    \caption{Coverage results of C@k metric.}
    \label{fig:Mattheweffect}
\end{figure}

\begin{table}
\small
\setlength{\tabcolsep}{0.1mm}
\setlength{\abovecaptionskip}{2pt}  
\centering
\renewcommand{\arraystretch}{1.1}
\begin{tabular}{l@{\hskip 0.1in}
c@{\hskip 0.05in}c@{\hskip 0.05in}c@{\hskip 0in}
c@{\hskip 0.15in}
c@{\hskip 0.05in}c@{\hskip 0.05in}c@{\hskip 0in}
c@{\hskip 0.1in}}
\hline
\multirow{2}{*}{\textbf{Model}}&
\multicolumn{3}{c}{OpenDialKG}&
&
\multicolumn{3}{c}{DuRecDial}
&\\
\cline{2-4}
\cline{6-8}
\rule{0pt}{10pt}
&A@5&A@15&A@30&
&A@5&A@15&A@30&\\
\hline
KBRD&0.0025&0.0025&0.0088&
&0.0318&0.0562&0.0938&\\
KGSF&0.0051&0.0108&0.0182&
&0.0276&0.0534&0.0952&\\
ReDial&1.0000&0.9375&0.8333&
&1.0000&0.8824&0.9677&\\
TGReDial&0.0022&0.0043&0.0070&
&0.0137&0.0399&0.0796&\\
MHIM&0.0022&0.0044&0.0075&
&0.0228&0.0434&0.0789&\\
HiCore&\textbf{0.0017}&\textbf{0.0043}&\textbf{0.0065}&
&\textbf{0.0226}&\textbf{0.0423}&\textbf{0.0751}&\\
\hline
\multirow{2}{*}{\textbf{Model}}&
\multicolumn{3}{c}{OpenDialKG}&
&
\multicolumn{3}{c}{DuRecDial}
&\\
\cline{2-4}
\cline{6-8}
\rule{0pt}{10pt}
&L@5&L@15&L@30&
&L@5&L@15&L@30&\\
\hline
KBRD&0.2921&0.2782&0.2782&
&0.3758&0.4149&0.3406&\\
KGSF&0.2398&0.2482&0.3343&
&0.3314&0.4243&0.3302&\\
ReDial&1.0000&0.8750&0.8333&
&1.0000&0.8235&0.9677&\\
TGReDial&0.2737&0.2482&0.2757&
&0.3654&0.3803&0.3846&\\
MHIM&0.1919&0.2343&0.2617&
&0.3315&0.3488&0.2706&\\
\textbf{HiCore*}&\textbf{0.1906}&\textbf{0.2092}&\textbf{0.2343}&
&\textbf{0.3122}&\textbf{0.3267}&\textbf{0.2666}\\
\hline
\end{tabular}
\caption{\label{tab:Matthew_effect_evaluation} Results of Average Popularity (A@K) and Long Tail Ratio (L@K).}
\end{table}

Fig.\ref{fig:Mattheweffect} illustrates the experimental outcomes, showcasing the consistent superiority of HiCore in achieving the highest levels of \emph{Coverage} across all datasets in comparison to the most robust baselines. The heightened coverage metric highlights its exceptional ability to encompass a broad spectrum of the recommendation space by incorporating items from diverse categories. Additionally, as outlined in Table \ref{tab:Matthew_effect_evaluation}, our proposed method demonstrates the lowest values for \emph{Average Popularity} and \emph{Long Tail Ratio}. This evidence suggests that our method effectively mitigates the adverse effects of item popularity on recommendation outcomes and successfully addresses the long tail distribution of items. These results validate the effectiveness of our proposed approach in combating the Matthew effect in the CRS as users interact with the system over time, attributed to its capability to learn multi-level user interests through a series of hypergraphs from triple-channel setting, including group, joint, and purchase channels.

\begin{figure}[t]
    \centering
    \includegraphics[width=0.5\textwidth]{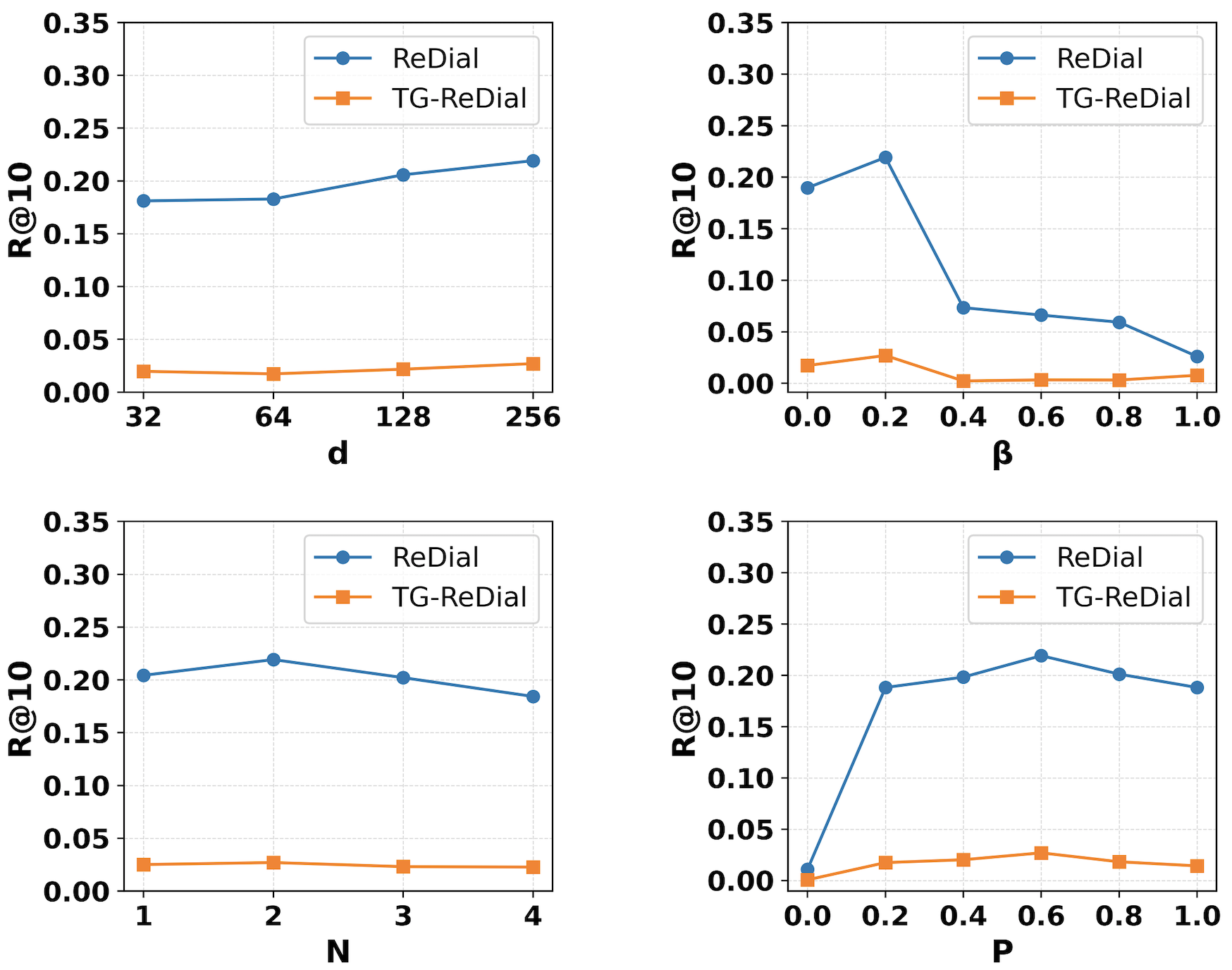}
    \caption{Impact of different hyperparameteres.}
    \label{fig:hyperparameteres}
\end{figure}

\subsection{Hyperparameters Analysis (RQ4)}
Hyperparameters are parameters in a machine learning algorithm that need to be manually set and tuned to optimize model performance, distinct from the parameters that the model learns during training. Next, we will delve into the research on how various hyperparameters influence the performance of recommendations, including the embedding dimension $d$, comparative learning weight $\beta$, hypergraph convolution layers $N$, and the hyperedge threshold $P$. From Fig.\ref{fig:hyperparameteres}, we can obtain: (1) Elevating the feature dimensionality enhances outcomes, as higher dimensions can encapsulate more intricate features effectively; (2) Having too few hyperedges may hinder the capture of intricate local patterns, whereas an excess of hyperedges could impede the model's convergence; (3) A lower beta value signifies a reduced weight for the comparison term, which show that the recommendation term exerts a more significant influence on the results; (4) A two-layer hyperconv network is sufficient to encode high-level features for enhancing recommendation performance.

\subsection{Ablation Studies (RQ5)}
To assess the efficacy of each component within the proposed method, we perform ablation experiments using various iterations of Hicore, including: 1) w/o $\textsf{G}^{i}_{g}$, w/o $\textsf{G}^{i}_{j}$, w/o $\textsf{G}^{i}_{p}$: removing item-oriented group-channel, joint-channel, purchase-channel hypergraph, respectively; 2) w/o $\textsf{G}^{e}_{g}$, w/o $\textsf{G}^{e}_{j}$, w/o $\textsf{G}^{i}_{p}$: removing entity-oriented group-channel, joint-channel, purchase-channel hypergraph, respectively; 3) w/o $\textsf{G}^{w}_{g}$, w/o $\textsf{G}^{w}_{j}$, w/o $\textsf{G}^{w}_{p}$: removing word-oriented group-channel, joint-channel, purchase-channel hypergraph, respectively.

Table \ref{tab:ablation_studies} outlines the experimental findings, indicating that the removal of any hypergraph type results in a performance decrease. This highlights the effectiveness of each hypergraph type and underscores the superiority of HiCore in learning multi-level user interests through a collection of hypergraphs to mitigate Matthew effect in the CRS.

\begin{table}
\small
\setlength{\tabcolsep}{5mm}
\setlength{\abovecaptionskip}{5pt}  
\centering
\begin{tabular}{l@{\hskip 0.1in}
c@{\hskip 0.05in}c@{\hskip 0in}
c@{\hskip 0.15in}
c@{\hskip 0.05in}c@{\hskip 0in}
c@{\hskip 0.1in}}
\toprule
\multirow{2}{*}{\textbf{Model}}&
\multicolumn{2}{c}{REDIAL}&
&
\multicolumn{2}{c}{TG-REDIAL}
&\\
\cline{2-3}
\cline{5-6}
\rule{0pt}{10pt}
&R@10&R@50&
&R@10&R@50&\\
\hline
\textbf{HiCore} &\textbf{0.2192}&\textbf{0.4163}&
&\textbf{0.0270}&\textbf{0.0769}\\
\hline
w/o $\textsf{G}^{(i)}_{g}$ &0.2075&0.4160&
&0.0234&0.0742\\
w/o $\textsf{G}^{(i)}_{j}$ &0.2012&0.4026&
&0.0217&0.0706\\
w/o $\textsf{G}^{(i)}_{p}$  &0.1939&0.4096&
&0.0220&0.0739\\
w/o $\textsf{G}^{(e)}_{g}$  &0.2067&0.4044&
&0.0247&0.0713\\
w/o $\textsf{G}^{(e)}_{j}$  &0.2142&0.4122&
&0.0253&0.0756\\
w/o $\textsf{G}^{(e)}_{p}$  &0.1971&0.4110&
&0.0243&0.0693\\
w/o $\textsf{G}^{(w)}_{g}$  &0.2067&0.4142&
&0.0264&0.0761\\
w/o $\textsf{G}^{(w)}_{j}$  &0.2151&0.4145&
&0.0223&0.0733\\
w/o $\textsf{G}^{(w)}_{p}$  &0.2067&0.3974&
&0.0263&0.0759\\

\bottomrule
\end{tabular}
\caption{\label{tab:ablation_studies} Ablation studies on the recommendation task.}
\end{table}

\section{Conclusion}
The Matthew effect poses a significant challenge in the CRS due to the dynamic user-system feedback loop, which tends to escalate over time as users engage with the system. In response to these challenges, we proposed a novel framework, HiCore, aimed at mitigating the Matthew effect by capturing multi-level user interests through a variety of hypergraphs, including item-oriented, entity-oriented, and word-oriented triple-channel hypergraphs. Extensive experiments validate that HiCore outperforms all baselines, demonstrating the effectiveness of HiCore in addressing the Matthew effect as users chat with the system over time in the CRS.

\section{Limitations}
While our HiCore has achieved a remarkable state-of-the-art performance, it does come with certain limitations. Firstly, triple-channel hypergraphs may present challenges due to their computational complexity, interpretational intricacies, and potential issues with sparse data. Secondly, scaling these hypergraphs to larger datasets could introduce scalability hurdles, with a risk of overfitting when the model becomes excessively fine-tuned to the training data. Furthermore, ensuring generalizability and handling resource-intensive computations are crucial factors to consider when leveraging multi-channel hypergraphs.

\section{Ethics Statement}
The data used in this paper are sourced from open-access repositories, and do not pose any privacy concerns. We are confident that our research adheres to the ethical standards set forth by EMNLP.

\section{Acknowledgements}
This research / project is supported by the National Research Foundation, Singapore and Infocomm Media Development Authority under its Trust Tech Funding Initiative. Any opinions, findings and conclusions or recommendations expressed in this material are those of the author(s) and do not reflect the views of National Research Foundation, Singapore and Infocomm Media Development Authority.

\bibliography{anthology,custom}
\bibliographystyle{acl_natbib}

\appendix

\end{document}